\documentclass{sig-alternate}
\usepackage{algorithm}
\usepackage[noend]{algorithmic}
\usepackage{url}
\usepackage{wrapfig}
\usepackage[usenames]{color}
\usepackage{graphicx}
\usepackage{epstopdf}
\usepackage{xcolor,colortbl}

\usepackage{color}
\usepackage{float}
\usepackage{fancyvrb}
\usepackage{multirow}
\usepackage{hhline}

\def\riq{$\mathsf{RIQ}$}

\newtheorem{theorem}{Theorem}

\newtheorem{definition}{Definition}

\long\def\comment#1{}
\definecolor{Gray}{gray}{0.95}

\newcommand{\linespace}[1]
        {\renewcommand{\baselinestretch}{#1}\Large\normalsize}
\newcommand{\algolinespace}{0.8}
\newcommand{\textlinespace}{1.0}

\begin{document}

\title{Fast Processing of SPARQL Queries on RDF Quadruples}
\numberofauthors{5}
\author{%
\alignauthor
Vasil Slavov \\
\affaddr{Univ. of Missouri-Kansas City} \\
\email{vgslavov@mail.umkc.edu}
\alignauthor
Anas Katib \\
\affaddr{Univ. of Missouri-Kansas City} \\
\email{anaskatib@mail.umkc.edu}
\alignauthor 
Praveen Rao \thanks{\footnotesize{\textcolor{blue}{An extended version of this WebDB 2014 paper has been published in the Journal of Web Semantics (JWS). (DOI: http://dx.doi.org/10.1016/j.websem.2016.03.005)}}} \\
\affaddr{Univ. of Missouri-Kansas City} \\
\email{raopr@mail.umkc.edu}
\and
\alignauthor 
Srivenu Paturi \\
\affaddr{Univ. of Missouri-Kansas City} \\
\email{sp895@mail.umkc.edu}
\alignauthor 
Dinesh Barenkala \\
\affaddr{Univ. of Missouri-Kansas City} \\
\email{db985@mail.umkc.edu}
}%

\maketitle

\newcommand{\ie}{{\em i.e.}}
\newcommand{\eg}{{\em e.g.}}
\newcommand{\etal}{{\em et al.\mbox{$\:$}}}
\newcommand{\noexists}{\mbox{$\backslash\!\!\!\!\!\;\exists$}}
\newcommand{\BigO}[1]{\mbox{${\cal O}(#1)$}}
\newcommand{\BigOmega}[1]{\mbox{$\Omega(#1)$}}
\newcommand{\BigTheta}[1]{\mbox{$\Theta(#1)$}}
\newcommand{\horizbar}{\rule{\linewidth}{.5mm}}
\newcommand{\ceiling}[1]{\left\lceil #1 \right\rceil}
\newcommand{\floor}[1]{\left\lfloor #1 \right\rfloor}
\newcommand{\faM}{\lfloor \alpha M \rfloor}
\newcommand{\C}[2]{{#1 \choose #2}}
\newcommand{\cardinal}[1]{\mbox{$\,\mid\!\! #1 \!\!\mid\,$}}
\newcommand{\xor}{\oplus}
\newcommand{\abs}[1]{\mbox{$\left|#1\right|$}}

\newcommand{\lpnorm}[2]{\mbox{$\left\|#1\right\|_{#2}$}}
\newcommand{\ltwonorm}[1]{\mbox{$\left\|#1\right\|_{2}$}}
\newcommand{\nrom}[1]{\lVert#1\rVert}
\newcommand{\lland}{\,\wedge\,}
\newcommand{\llor}{\,\vee\,}
\newcommand{\xby}{\!\times\!}
\newcommand{\combi}[2]{\mbox{$\left(\!\begin{array}{c}{#1}\\{#2}\end{array}\!\right)$}}
\newcommand{\tilda}{\verb+~+}
\newcommand{\pair}[1]{\mbox{$\langle #1 \rangle$}}
\newcommand{\edge}[1]{\mbox{$\overline{#1}$}}

\def\select#1#2{\mbox{$\sigma_{#1}(#2)$}}

\newcommand{\openbox}{\leavevmode
  \hbox to.77778em{%
  \hfil\vrule
  \vbox to.675em{\hrule width.6em\vfil\hrule}%
  \vrule\hfil}}
\newcommand{\qedsymbol}{\openbox}
\newcommand{\proofbegin}{{\em Proof. \ }}               
\newcommand{\proofend}{\hfill\qedsymbol\bigskip}        

\begin{abstract}
In this paper, we propose a new approach for fast processing of SPARQL
queries on large RDF datasets containing RDF quadruples (or
quads). Our approach called \riq{} employs a {\em
  decrease-and-conquer} strategy: Rather than indexing the entire RDF
dataset, \riq{} identifies groups of similar RDF graphs and indexes
each group separately. During query processing, \riq{} uses a novel
filtering index to first identify candidate groups that may contain
matches for the query. On these candidates, it executes optimized
queries using a conventional SPARQL processor to produce the final
results. Our initial performance evaluation results are promising:
Using a synthetic and a real dataset, each containing about 1.4
billion quads, we show that \riq{} outperforms RDF-3X and Jena TDB on
a variety of SPARQL queries.
\end{abstract}

\section{Introduction}

The Resource Description Framework (RDF) is a standard model for
representing data on the Web~\cite{RDF}. It enables the interchange
and machine processing of data by considering its semantics. While RDF
was first proposed with the vision of enabling the Semantic Web, it
has now become popular in domain-specific applications and the
Web. Through advanced RDF technologies, one can perform semantic
reasoning over data and extract knowledge in domains such as
healthcare, biopharmaceuticals, defense, and intelligence. Linked
Data~\cite{LinkedData09} is a popular use case of RDF on the Web; it
has a large collection of different knowledge bases, which are
represented in RDF (\eg, DBpedia~\cite{DBPedia}).

With a growing number of new applications relying on Semantic Web
technologies (\eg, Pfizer~\cite{Pfizer}; Newsweek, BBC, The New York
Times, and Best Buy~\cite{MITERMIT}) and the availability of large RDF
datasets (\eg, Billion Triples Challenge
(BTC)~\cite{BillionTriplesChallenge}, Linking Open Government Data
(LOGD)~\cite{LOGD}), there is a need to advance the state-of-the-art
in storing, indexing, and query processing of RDF datasets.



Today, datasets containing over a billion RDF quads are becoming
popular on the Web (\eg, BTC~\cite{BillionTriplesChallenge},
LOGD~\cite{LOGD}). Such datasets can be viewed as a collection of RDF
graphs. Using SPARQL's {\tt GRAPH} keyword~\cite{SPARQL1.1}, one can
pose a query to match a specific graph pattern within any single RDF
graph. While researchers in the database community have proposed
scalable approaches for indexing and query processing of large RDF
datasets~\cite{Abadi2009SVP,Hexastore,RDF-3X10,BitMat,Abadi2011,Bornea2013,Yuan2013,Zeng2013},
they have designed these techniques for RDF datasets containing
triples. In addition, none of them have investigated how large and
complex graph patterns in SPARQL queries can be processed
efficiently. Evidently, RDF-3X~\cite{RDF-3X10}, a popular scalable
approach for a local/centralized environment, yields poor performance
when SPARQL queries containing large graph patterns are processed over
large RDF datasets. This is because of the large number of join
operations that must be performed to process a query.


We posit that, on RDF datasets containing billions of quads, any
approach that first finds matches for subpatterns in a large graph
pattern and then employs join operations to merge partial matches will
face a similar limitation. Motivated by the aforementioned reasons, we
propose a new approach called \riq{} (\textbf{R}DF \textbf{I}ndexing
on \textbf{Q}uads) and make the following contributions in this paper:

\textbullet~ We propose a new vector representation for RDF graphs and
graph patterns in SPARQL queries. This representation enables us to
group similar RDF graphs and index each group separately rather than
constructing an index on the entire dataset. We propose a novel
filtering index, which employs a combination of Bloom Filters and
Counting Bloom Filters to compactly store it.

\textbullet~ We propose a {\em decrease-and-conquer} approach to
efficiently process a SPARQL query. Using the filtering index, we can
methodically and quickly identify candidate groups of RDF graphs that
may contain a match for the query. We can then execute optimized
queries on the candidates using a conventional SPARQL processor that
supports quads.

\textbullet~ We report the results from our initial performance
evaluation using a synthetic and a real dataset, each containing about
1.4 billion quads. We observed that \riq{} can outperform RDF-3X and
Jena TDB on a variety of SPARQL queries.

\vspace*{-2ex}
\section{Background}

The RDF data model provides a simple way to represent any assertion as
a (subject, predicate, object) triple. A collection of triples can be
modeled as a directed, labeled graph. If each triple has a graph name
(or context), it is called a quad. Below is an example of a quad from
the BTC 2012 dataset~\cite{BillionTriplesChallenge} with its subject,
predicate, object, and context:
\url{<http://data.linkedmdb.org/resource/producer/10138>}
\url{<http://data.linkedmdb.org/resource/movie/producer_name>} ``{\tt
  Mani Ratnam}" \url{<http://data.linkedmdb.org/data/producer/10138>}.
Triples with the same context belong to the same RDF graph.

\begin{figure}
\begin{scriptsize}
\begin{Verbatim}[frame=single]
PREFIX movie: <http://data.linkedmdb.org/resource/movie/>
PREFIX rdfs: <http://www.w3.org/2000/01/rdf-schema#>
PREFIX foaf: <http://xmlns.com/foaf/0.1/>
SELECT ?g ?producer ?name ?label ?page ?film WHERE {
  GRAPH ?g { ?producer movie:producer_name ?name .
             ?producer rdfs:label ?label .
             OPTIONAL { ?producer foaf:page ?page . } .
             ?film movie:producer ?producer .  } }
\end{Verbatim}
\end{scriptsize}
\vspace*{-3ex}
\caption{An example of a SPARQL query}
\label{fig-example1}
\vspace*{-3ex}
\end{figure}

Using SPARQL, one can express complex graph pattern queries on RDF
graphs. One of the fundamental operations in RDF query processing is
{\em Basic Graph Pattern Matching}~\cite{SPARQL1.1}. A Basic Graph
Pattern (BGP) in a query combines a set of triple patterns. A triple
pattern contains variables (prefixed by {\tt ?})  and constants.
During query processing, the variables in a BGP are bound to RDF terms
in the data, \ie, the nodes in the same RDF graph, via subgraph
matching~\cite{SPARQL1.1}. Common variables within a BGP or across
BGPs denote a join operation on the variable bindings of triple
patterns. Consider the query shown in Figure~\ref{fig-example1}. The
bindings for the subject (variable) in {\tt ?producer
  movie:producer\_name ?name} are joined with the bindings for the
object (variable) in {\tt ?film movie:producer ?producer}. The
variable {\tt ?g} will be bound to the names/contexts of those RDF
graphs that contain a match for the graph pattern specified inside the
{\tt GRAPH} block. {\tt OPTIONAL} allows certain patterns to have
empty bindings; {\tt UNION} combines bindings of multiple graph
patterns.

\section{Related Work and Motivation}
Several approaches have been developed for indexing and querying RDF
data in a local/centralized environment. Early approaches employed an
RDBMS to store and query RDF data (\eg, Sesame~\cite{Sesame},
Oracle~\cite{OracleRDF05}). Unfortunately, the cost of self-joins on a
single (triples) table became a serious bottleneck. Later, Abadi~\etal
proposed the idea of vertically partitioning the property
tables~\cite{Jena06} and used a column-oriented DBMS to achieve an
order of magnitude performance improvement over previous
techniques~\cite{Abadi2009SVP}. Recently, Neumann \etal developed
RDF-3X~\cite{RDF-3X10} that builds exhaustive indexes on the six
permutations of $(s,p,o)$ triples. RDF-3X significantly outperformed
the vertical partitioning approach. It uses a new join ordering method
based on selectivity estimates and builds compressed
indexes. Weiss~\etal~\cite{Hexastore} developed Hexastore that also
builds exhaustive indexes. However, Hexastore suffers from large index
sizes due to lack of compression. Atre~\etal~\cite{BitMat} developed
BitMat to overcome the overhead of large intermediate join results for
queries containing low selectivity triple patterns. BitMat performs
in-memory processing of compressed bit matrices during query
processing.

More recently, Bornea~\etal~\cite{Bornea2013} developed DB2RDF by
using an RDBMS to store and query RDF data. By storing the
predicate-object pairs of each subject in the same row of the
relational table, they reduced the number of joins required for
star-shaped BGPs. DB2RDF maintains only subject and object indexes and
employs a novel SPARQL-to-SQL translation technique for generating
optimized queries. Yuan~\etal~\cite{Yuan2013} developed TripleBit,
which uses a compact storage scheme for RDF data by representing
triples via a Triple Matrix. For each predicate, TripleBit maintains
SO and OS ordered buckets. Using a collection of indexes and optimal
join ordering, it reduces the size of the intermediate results during
query processing.

A few approaches exploit the graph properties/structure of RDF data
for indexing and query
processing~\cite{Sintek2006,Udrea07,DOGMA09,Zou2011,PicalausaLFHV12}. These
techniques, however, have been tested only on small RDF datasets
containing less than 50 million triples. Recently, a few schemes were
proposed for distributed/parallel RDF query
processing~\cite{Abadi2011,Zeng2013}. Our work, however, focuses on
RDF query processing in a local environment.

The motivation for our work stems from two key observations: First,
the above approaches were designed to process RDF datasets containing
triples. Simply ignoring the context in an RDF quad and using an
existing approach designed for triples may produce incorrect results
due to bindings for a BGP from different
graphs~\cite{tech-report2014}. Second, most of the queries tested by
these approaches contain BGPs with a modest number of triples patterns
(at most 8). None of them have investigated how to efficiently process
SPARQL queries with large, complex BGPs (\eg, containing undirected
cycles\footnote{Here is a simple example: \{?a p ?b . ?b q ?c . ?a r ?c .\}.}).

\section{The Design of \riq}

\begin{figure}
\begin{scriptsize}
\begin{Verbatim}[frame=single]
Query => 'SELECT' Variables 'WHERE' '{' 'GRAPH' Variables 
  '{' GroupGraphPattern '}' '}' ResultModifiers
GroupGraphPattern => BGP? ( GraphPatternNotTriples '.'? BGP? )*
GraphPatternNotTriples => 
  GroupOrUnionGraphPattern | OptionalGraphPattern | Filter
GroupOrUnionGraphPattern => 
  GroupGraphPattern ( 'UNION' GroupGraphPattern )*
OptionalGraphPattern => 'OPTIONAL' GroupGraphPattern
Filter => 'FILTER' Constraint 
Constraint => Predicate | 'EXISTS' BGP | 'NOT EXISTS' BGP
\end{Verbatim}
\end{scriptsize}
\vspace*{-3ex}
\caption{Grammar for queries}
\label{fig-grammar}
\vspace*{-2ex}
\end{figure}

\begin{figure}
\begin{center}
\includegraphics*[width=2.8in, angle=0]{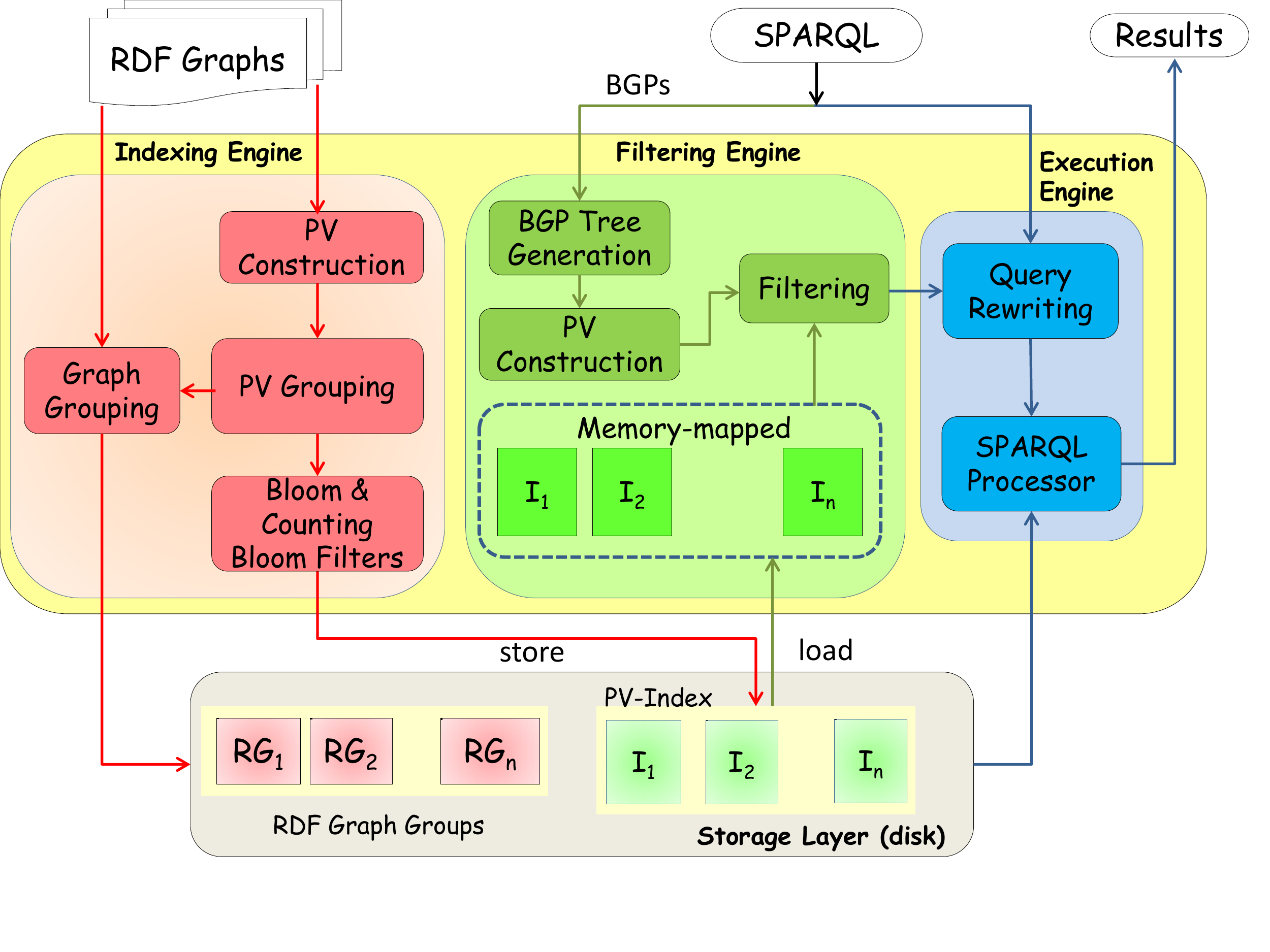}
\end{center}
\vspace*{-9ex}
\caption{Overview of \riq}
\label{fig-system-overview}
\vspace*{-2ex}
\end{figure}

In this section, we present the design of \riq{} (RDF Indexing on
Quadruples) and describe its three main components, \ie, the Indexing
Engine, the Filtering Engine, and the Execution Engine. (See
Figure~\ref{fig-system-overview}.) Our goal is to support a subset of
the SPARQL grammar~\cite{SPARQL1.1} as shown in
Figure~\ref{fig-grammar}.

\subsection{Indexing RDF Data}
\label{sec-indexingengine}

We introduce a new vector representation for RDF graphs and BGPs,
which will allows us to capture the properties of the triples and
triple patterns in them. This vector representation plays a key role
in the construction of an effective filtering index, where similar RDF
graphs will be grouped together.


\subsubsection{Essential Transformations}
To begin with, we define two transformations: one for a triple in an
RDF graph and the other for a triple pattern in a BGP. Let
$\mathbb{P}$ = \{$SPO$, $SP?$, $S?O$, $?PO$, $S??$, $?P?$, $??O$\} be
a set of canonical patterns.  We denote the transformation on a triple
(s,p,o) by $f_D: \mathbb{P} \times \{(s,p,o)\} \rightarrow O_D$, where
the range $O_D$ is shown Table~\ref{table-transformations} for each
canonical pattern. Note that $O_D$ resembles triple patterns (variable
names excluded) that can appear in a BGP.

\begin{table}
\begin{tabular}{|l||l|}
\hline
\multicolumn{1}{|c||}{Transformation $f_D$} &
\multicolumn{1}{|c|}{Transformation $f_Q$} \\
\hline
\hline
{$f_D$\small{(SPO, (s,p,o))} = {(s,p,o)}} & {$f_Q$(`s p o') = {\small{(SPO,(s,p,o))}}} \\
\hline
{$f_D$\small{(SP?, (s,p,o))} = {(s,p,?)}} & {$f_Q$(`s p ?$v_o$') = {\small{(SP?,(s,p,?))}}} \\
\hline
{$f_D$\small{(S?O, (s,p,o))} = {(s,?,o)}} & {$f_Q$(`s ?$v_p$ o') = {\small{(S?O,(s,?,o))}}} \\
\hline
{$f_D$\small{(?PO, (s,p,o))} = {(?,p,o)}} & {$f_Q$(`?$v_s$ p o') = {\small{(?PO,(?,p,o))}}} \\
\hline
{$f_D$\small{(S??, (s,p,o))} = {(s,?,?)}} & {$f_Q$(`s ?$v_p$ ?o') = {\small{(S??,(s,?,?))}}} \\
\hline
{$f_D$\small{(?P?, (s,p,o))} = {(?,p,?)}} & {$f_Q$(`?$v_s$ p ?$v_o$') = {\small{(?P?,(?,p,?))}}} \\
\hline
{$f_D$\small{(??O, (s,p,o))} = {(?,?,o)}} & {$f_Q$(`?$v_s$ ?$v_p$ o') = {\small{(??O,(?,?,o))}}} \\
\hline
\end{tabular}
\vspace*{-2ex}
\caption{Transformations in \riq}
\label{table-transformations}
\vspace*{-2ex}
\end{table}


Next, we denote a transformation $f_{Q}: T \rightarrow \mathbb{P}
\times O_{Q}$, where $T$ denotes the set of triple patterns that can
appear in a query. The range $\mathbb{P} \times O_{Q}$ is shown in
Table~\ref{table-transformations} and identifies the canonical pattern
for a given triple pattern. Although the triple pattern `s p o' has no
variables, it is still a valid triple pattern in a
BGP.\footnote{SELECT ?g WHERE \{ GRAPH ?g \{ s p o . \} \}.}

The transformations $f_{D}$ and $f_{Q}$ allow us to map a triple in
the data and a triple pattern in a query to a common plane of
reference. This will enable us to quickly test {\em if a triple
  pattern in a BGP has a match in the data}.

\subsubsection{Pattern Vectors}

Given an RDF graph with context $c$, we map it into a vector
representation called a Pattern Vector (PV) and denote it by
$\overline{V_c}$. Essentially, $\overline{V_c}$ = ($V_{c,SPO}$,
$V_{c,SP?}$, $V_{c,S?O}$, $V_{c,?PO}$, $V_{c,S??}$, $V_{c,?P?}$,
$V_{c,??O}$), where each $V_{c,r}$ denotes the vector constructed for
$r \in \mathbb{P}$. We assume a hash function $\mathbb{H}: B
\rightarrow \mathbb{Z}^*$, where $B$ denotes a bit string and the
range is the set of non-negative integers. Now, we construct
$\overline{V_{c}}$ as follows: Initially, each $V_{c,r}$ is
empty. Given a quad $(s,p,o,c)$ in the graph, for each $r \in
\mathbb{P}$, we compute $\mathbb{H}(f_D(r,(s,p,o)))$ and insert it
into $V_{c,r}$. We perform this computation on every quad in the graph
to generate $\overline{V_c}$. Note that $\overline{V_c}$ requires
space linear in the number of quads in the graph.


Our hash function $\mathbb{H}$ is based on Rabin's fingerprinting
technique~\cite{tech-reportRabin}, which is efficient to compute. If
we generate 32-bit hash values, the probability of collision is
extremely low~\cite{tech-report2014}. Thus, in practice, we can view
$V_{c,SPO}$ as a set, because the quads/triples in a graph are always
assumed to be unique. However, the remaining vectors of
$\overline{V_c}$ should be viewed as multisets, because $f_{D}$ can
produce the same output for different triples due to the presence of
`?' in the output.



Given a BGP $q$, we map it into a PV, denoted by $\overline{V_q}$, and
compute it slightly differently: Initially, each $V_{q,r}$ is
empty. For each triple pattern $t$ in $q$, we compute $f_Q(t)$ to
produce a pair $(r,o)$, where $r$ denotes the canonical pattern for
$t$. We then insert $\mathbb{H}(o)$ into $V_{q,r}$. As before,
$V_{q,SPO}$ can be viewed as a set. The rest of the vectors of
$\overline{V_{q}}$ should be viewed as multisets, because two
different triple patterns (each containing at least one variable) in a
BGP may hash to the same value. For example, if a BGP contains two
triple patterns {\tt ?$s_1$ movie:producer ?$o_1$} and {\tt ?$s_2$
  movie:producer ?$o_2$}, then $f_Q$(`{\tt $?s_1$ movie:producer
  $?o_1$}') = $f_Q$(`{\tt $?s_2$ movie:producer $?o_2$}') and
therefore, the hash values produced by $\mathbb{H}$ will be identical.


\subsubsection{Operations on Pattern Vectors}

Next, we define two operations on PVs, which will be used during the
construction of the filtering index. Our goal is to group similar PVs
(and as a result, similar RDF graphs) together so that candidate RDF
graphs are identified and processed quickly during query processing.

\begin{definition}[Union]
Given two PVs, say $\overline{V_a}$ and $\overline{V_b}$, their union
$\overline{V_a} \cup \overline{V_b}$ is a PV say $\overline{V_{c}}$,
where $V_{c,r} \leftarrow V_{a,r} \cup V_{b,r}$ and $r \in
\mathbb{P}$.
\end{definition}
\begin{definition}[Similarity]
Given two PVs, say $\overline{V_a}$ and $\overline{V_b}$, their
similarity is denoted by $sim(\overline{V_{a}},\overline{V_{b}})$ =
$\max \limits_{r\in\mathbb{P}}$ $sim(V_{a,r}, V_{b,r})$, where
$sim(V_{a,r},V_{b,r}) = \frac{|V_{a,r} \cap V_{b,r}|}{|V_{a,r} \cup
  V_{b,r}|}$.
\end{definition}


\subsubsection{Index Construction}

We begin by describing a key necessary condition, which forms the
basis for indexing and query processing in \riq.  Because we map both
the RDF graphs and BGPs into their PVs, we must characterize the
relationship between them when processing a BGP -- assuming it is a
connected graph -- via subgraph matching. We state the following
theorem.

\begin{theorem}
\label{thrm-bgpmatch}
Suppose $\overline{V_c}$ and $\overline{V_q}$ denote the PVs of an RDF
graph and a BGP, respectively. If the BGP has a subgraph match in the
RDF graph, then $\bigwedge \limits_{r \in \mathbb{P}} (V_{q,r}
\subseteq V_{c,r}) = $ {\tt TRUE}. (See the technical
report~\cite{tech-report2014} for the proof.)
\end{theorem} 

According to Theorem~\ref{thrm-bgpmatch}, given a BGP, if we can
identify those RDF graphs in the database whose PVs satisfy the
necessary condition, then we have a superset of RDF graphs that
contain a subgraph match for the BGP. This also guarantees that there
are no false dismissals.


Rather than testing every PV in the database -- one-at-a-time --
during query processing, we propose a novel filtering index called the
PV-Index to effectively organize millions of PVs in the
database. Using this index, we aim to quickly identify candidate RDF
graphs in the early stages of query processing using
Theorem~\ref{thrm-bgpmatch}. Our goal is to discard most of the
non-matching RDF graphs without any false dismissals. As a result, the
subsequent stages of query processing will process fewer candidates to
obtain the final results, thereby speeding up query processing.



There are two issues that arise while designing the PV-Index: First,
we want to group similar PVs together so that for a given BGP, we can
quickly discard most of the non-matching RDF graphs. Second, we want
to compactly store the PV-Index to minimize the cost of I/O during
query processing. To address the first issue, we use the concept of
locality sensitive hashing (LSH)~\cite{lsh98}. For similarity on sets
based on the Jaccard index, LSH on a set $S$, denoted by
$\mathsf{LSH}_{k,l,m}(S)$ can be performed as
follows~\cite{Haveliwala2002}: Pick $k \times l$ random linear hash
functions of the form $h(x) = (ax + b)~mod~u$, where $u$ is a prime,
and $a$ and $b$ are integers such that $0< a < u$ and $0 \leq b <
u$. Compute $g(S) = \min \{h(x)\}$ over all items in the set as the
output hash value for $S$.  Each group of $l$ hash values is hashed
(\eg, using Rabin's fingerprinting) to the range $[0,m-1]$. This
results in $k$ hash values for $S$. It is known that given two sets
$S_1$ and $S_2$ with similarity $p = \frac{|S_1 \cap S_2|}{|S_1 \cup
  S_2|}$, $\Pr [g(S_1) = g(S_2)] = p$. Also, the probability that
$\mathsf{LSH}_{k,l,m}(S_1)$ and $\mathsf{LSH}_{k,l,m}(S_2)$ have at
least one hash value identical is $1 - (1 - p^{l})^{k}$. The above
properties also hold for multisets.

To address the second issue, we employ Bloom filters (BFs) and
Counting Bloom filters (CBFs)~\cite{broder2003} to compactly represent
the PV-Index. A Bloom filter is a popular data structure to compactly
represent a set of items and process membership queries on it. A
Counting Bloom filter maintains $n$-bit counters instead of single
bits and can represent multisets. Both BFs and CBFs can be configured
to achieve a false positive rate based on their
capacities~\cite{broder2003}.


\linespace{\algolinespace}

\begin{algorithm}
\caption{The PV-Index Construction}
\label{algo-index-construction}
\begin{algorithmic}[1]
\REQUIRE a list of PVs; $(k,l,m)$: LSH parameters; $\epsilon$: false positive rate
\ENSURE filters of all the groups of similar RDF graphs
\STATE Let $\mathbb{G}(\mathbb{V},\mathbb{E})$ be initialized to an empty graph
\FOR{each PV $\overline{V}$} \label{line-create}
\STATE Add a new vertex $v_i$ to $\mathbb{V}$
\FOR{each $r \in \mathbb{P}$}
\STATE $\{h_{i1},...,h_{ik}\} \leftarrow \mathsf{LSH}_{k,l,m}(V_r)$
\FOR{every $v_j \in \mathbb{V}$ and $i\neq j$}
\IF{$\exists~o$ s.t. $1 \leq o \leq k$ and $h_{io} = h_{jo}$}
\STATE Add an edge $(v_i,v_j)$ to $\mathbb{E}$ if not already present 
\ENDIF\label{line-create-end}
\ENDFOR
\ENDFOR
\ENDFOR
\STATE Compute the connected components of $\mathbb{G}$. Let $\{C_1,...,C_t\}$ denote these components.
\FOR{$i=1$ to $t$}
\STATE Compute the union $U_i$ of all PVs corresponding to the vertices in $C_i$ \label{line-union}
\STATE Construct a BF for $U_{i,SPO}$ with false positive rate $\epsilon$ given the capacity $|U_{i,SPO}|$ \label{line-BF}
\STATE Construct a CBF for each of the remaining vectors of $U_i$ with false positive rate $\epsilon$ given the capacity $|U_{i,*}|$ \label{line-CBF}
\STATE Store the ids of graphs belonging to $C_i$
\ENDFOR
\RETURN 
\end{algorithmic}
\end{algorithm}
\linespace{\textlinespace}

In Algorithm~\ref{algo-index-construction}, we outline the steps to
construct the PV-Index. We build a graph $\mathbb{G}$, where each
vertex of $\mathbb{G}$ represents a PV. For every PV, we apply LSH on
each of its seven vectors. Suppose there are two PVs such that the
application of LSH on their vectors for the same pattern $r$, produces
at least one identical hash value, then we add an edge between the
vertices representing these PVs (Lines~\ref{line-create}
to~\ref{line-create-end}). Essentially, a missing edge between two
vertices indicates that their corresponding PVs are dissimilar with
high probability. Once $\mathbb{G}$ is constructed, we compute (in
linear time) the connected components in it. Each connected component
represents RDF graphs whose corresponding PVs are similar with high
probability. We treat these graphs as a group and compute the union of
their PVs (Line~\ref{line-union}). The union operation summarizes the
PVs as well as preserves the condition stated in
Theorem~\ref{thrm-bgpmatch}. (The individual vectors in a PV are
sorted to enable the union operation in linear time.)


To compactly represent the union computed for a connected component,
we use a combination of one Bloom filter (BF) and six Counting Bloom
filters (CBFs). The vector for the canonical pattern SPO is stored
using a BF and the others are stored using CBFs. Each filter of a
vector is configured for a false positive rate of $\epsilon$ and
capacity equal to the cardinality of the vector (Lines~\ref{line-BF}
and~\ref{line-CBF}).  For each connected component, we also store the
ids of graphs belonging to it. In summary, the BFs and CBFs for all
the connected components constitute the PV-Index. Each group of graphs
is separately indexed using a tool like Jena TDB.

\subsection{Query Processing}
\label{sec-queryprocessing}
Next, we propose a {\em decrease-and-conquer} approach for efficient
SPARQL query processing in \riq. That is, we first identify candidate
groups of RDF graphs that may contain matches for a query using the
PV-Index and then execute optimized SPARQL queries on these
candidates.

\begin{figure}[tb]
\begin{center}
\includegraphics*[width=3.2in, angle=0]{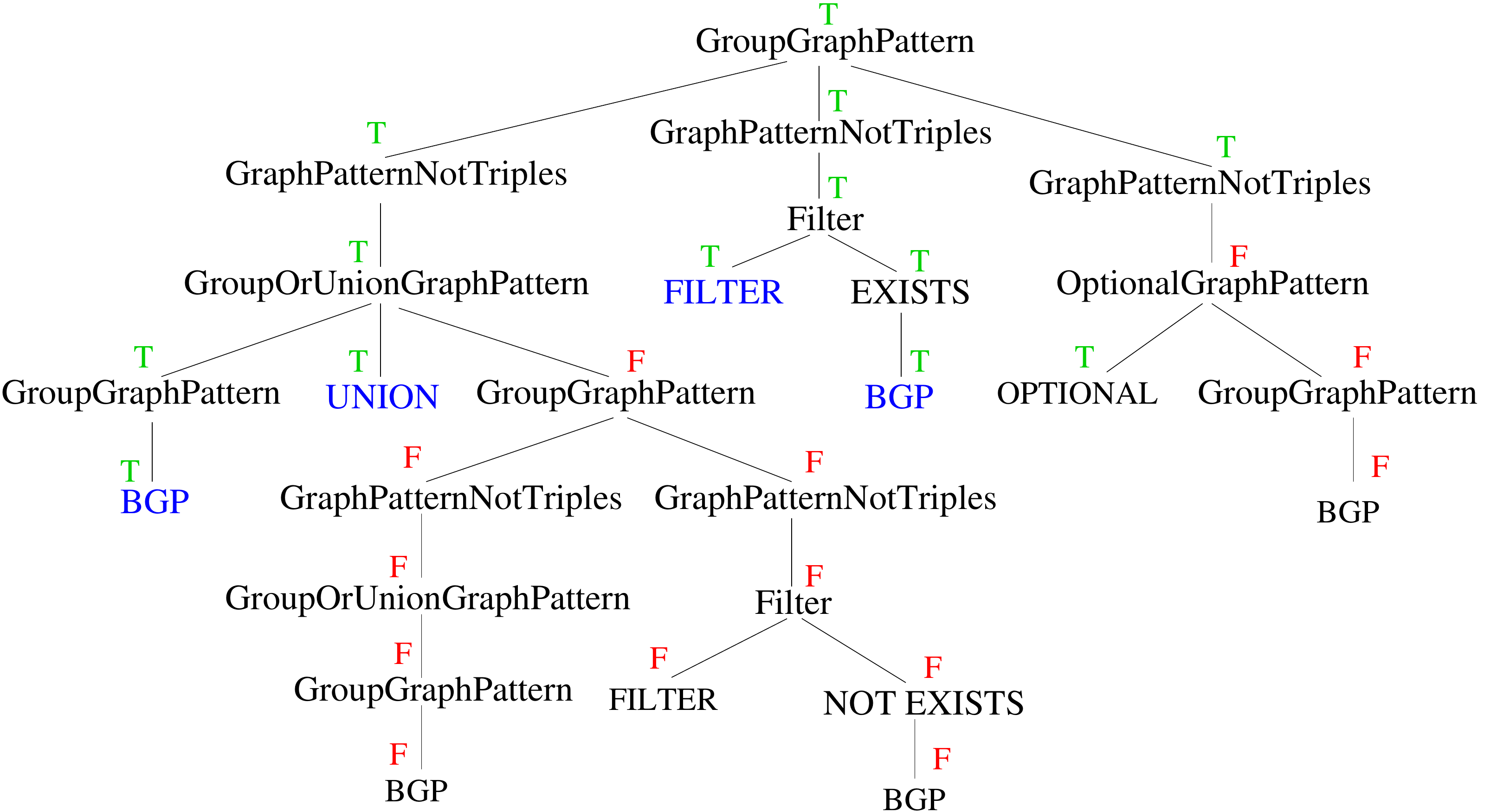}
\vspace*{-2ex}
\caption{An example of a BGP Tree}
\label{fig-BGPTree}
\vspace*{-4ex}
\end{center}
\end{figure}

Given a query, the first step is to parse its {\tt GRAPH} block
according to the grammar in Figure~\ref{fig-grammar} and generate a
tree-representation, which we call the BGP Tree. This tree serves as
an execution plan for processing individual BGPs in the query. (See
Figure~\ref{fig-BGPTree} for an example.)  We maintain a Boolean
variable $eval[n]$ for each node $n$ in the tree to denote the status
of the evaluation on a connected component of the PV-Index. With
$eval[n]=$ {\tt FALSE} for every node in the tree, we invoke
Algorithm~\ref{algo-bgptree-execution} on each connected component,
starting from the root of the BGP Tree. When a child of {\tt
  GroupGraphPattern} evaluates to {\tt FALSE}, we skip processing the
remaining children (Line~\ref{line-eval-skipnodes}), because the RDF
graphs belonging to that connected component will not produce a match
for the subexpression rooted at {\tt GroupGraphPattern}. For {\tt
  GroupOrUnionGraphPattern}, however, at least one of its children
\ie, {\tt GroupGraphPattern}, should evaluate to {\tt TRUE} to produce
a match (Line~\ref{line-eval-union}).

When a BGP is encountered (Line~\ref{line-eval-filter}), we test the
necessary condition stated in Theorem~\ref{thrm-bgpmatch} by calling
Algorithm~\ref{algo-bgpmatch}. This involves the processing of
membership queries on the BF and CBFs constructed for that connected
component. If {\tt OptionalGraphPattern} evaluates to {\tt FALSE}, we
return {\tt TRUE} because of the semantics of {\tt OPTIONAL} in
SPARQL. If $eval[root]$ = {\tt TRUE}, then the group of RDF graphs
belonging to that connected component is a candidate for further
processing.

For the candidate, an optimized SPARQL query can be generated by
traversing the BGP Tree and checking the evaluation status of each
node. (In the interest of space, we provide the algorithm in the
technical report~\cite{tech-report2014}.)

The result modifiers and predicates within {\tt FILTER} are included
in the optimized query.  In Figure~\ref{fig-BGPTree}, we show an
example, where the {\tt OPTIONAL} block and one block in the {\tt
  UNION} will be discarded in the optimized query. The optimized query
can then be executed on the candidate using a tool like Jena TDB. The
results from all the candidates should be combined to produce the final
results.

\linespace{\algolinespace}
\begin{algorithm}[tb]
\caption{EvalBGPTree(node $n$, conn. component $j$)}
\label{algo-bgptree-execution}
\begin{algorithmic}[1]
\STATE Let $c_1,...,c_{\tau}$ denote the child nodes of $n$ (left-to-right)
\FOR{$i$ = $1$ to $\tau$}
\STATE $eval[c_{i}]$ $\leftarrow$ EvalBGPTree($c_{i}$, $j$)
\IF{$n$ is {\tt GroupGraphPattern} \& $eval[c_{i}]$ = {\tt FALSE}} \label{line-eval-skipnodes}
\STATE $eval[n]$ $\leftarrow$ {\tt FALSE}
\RETURN {\tt FALSE} \COMMENT{//skip rest of the nodes}
\ENDIF
\ENDFOR
\IF{$n$ is {\tt GroupOrUnionGraphPattern}} \label{line-eval-union}
\STATE $eval[n] \leftarrow \bigvee \limits_{i=1}^{\tau} eval[c_i]$ 
\ELSIF{$n$ is {\tt EXISTS}}
\STATE $eval[n] \leftarrow eval[c_1]$
\ELSIF{$n$ is {\tt NOT EXISTS}}
\STATE $eval[n] \leftarrow$ {\tt TRUE}
\ELSIF{$n$ is {\tt Predicate}}
\STATE $eval[n] \leftarrow$ {\tt TRUE} \COMMENT{//skip processing predicates}
\ELSIF{$n$ is {\tt BGP}}\label{line-eval-filter}
\STATE Let $q$ denote the basic graph pattern
\STATE $eval[n]$ $\leftarrow$ IsMatch$(q,j)$  
\ELSE
\STATE $eval[n] \leftarrow eval[c_{\tau}]$  
\ENDIF
\IF{$n$ is {\tt OptionalGraphPattern}}
\RETURN {\tt TRUE}
\ENDIF
\RETURN $eval[n]$
\end{algorithmic}
\end{algorithm}
\linespace{\textlinespace}

\linespace{\algolinespace}
\begin{algorithm}[tb]
\caption{IsMatch(BGP $q$, conn. component $j$)}
\label{algo-bgpmatch}
\begin{algorithmic}[1]
\STATE For connected component $j$, let $\mathbb{F}_{j,r}$ denote the BF or CBF constructed for pattern $r$
\STATE Construct $\mathbb{F}_{q,r}$ with the same capacity and false positive rate as $\mathbb{F}_{U_j,r}$
\IF{(1) for each bit in $\mathbb{F}_{q,SPO}$ set
to 1, the corresponding bit in $\mathbb{F}_{U_j,SPO}$ is 1, and (2)
for each of the remaining patterns, given a non-zero counter in
$\mathbb{F}_{q,r}$, the corresponding counter in $\mathbb{F}_{U_j,r}$
is greater than or equal to it}
\RETURN {\tt TRUE}, otherwise \textbf{return} {\tt FALSE}
\ENDIF
\end{algorithmic}
\end{algorithm}
\linespace{\textlinespace}

\section{Performance Evaluation}

In this section, we report the initial performance evaluation of
\riq{} and have compared it with the latest version of RDF-3X and
Apache Jena 2.11.1 (TDB). RDF-3X and Jena TDB readily index datasets
with more than a billion triples. Also, Jena supports RDF quads. We
ran all the experiments on a 64-bit Ubuntu 12.04 machine with 4 Intel
Xeon 2.4GHz cores and 16GB RAM. \riq{} uses popular
open-source libraries for parsing RDF data~\cite{raptor2} and
constructing BFs and CBFs~\cite{dablooms}. All the three approaches
were single-threaded.

\subsection{Datasets and Queries}

We used one synthetic and one real dataset in our experiments. The
synthetic dataset was generated using the Lehigh University Benchmark
(LUBM)~\cite{LUBM} and contained 1.38 billion triples, 18 unique
predicates, and 10,000 universities. The triples were divided across
200,004 files and each file was treated as one RDF graph. The real
dataset was BTC 2012~\cite{BillionTriplesChallenge}, which is widely
used in the Semantic Web community. It contained 1.36 billion RDF
quads with 57,000 unique predicates and 9.59 million RDF graphs.

For LUBM, the query set included 3 SPARQL queries with large, complex
BGPs (L1-L3) and 9 others (L4-L12) that are variations of the queries
in the LUBM benchmark. For BTC 2012, the query set also included 2
SPARQL queries with large, complex BGPs (B1, B2) and 5 others
(B3-B7). (In the interest of space, the queries are listed in the
technical report~\cite{tech-report2014}.) The number of triples
patterns in each query and the number of results obtained for each
query using the three approaches are shown in
Table~\ref{table-queries}.

\subsection{Index Size}

Here we report the size of the indexes built by the three
approaches. For LUBM, the size of the index built by RDF-3X and Jena
TDB were 77 GB and 121 GB, respectively. The filtering index of \riq{}
was 8.5 GB in size and had 339 unions. For BTC 2012, the size of the
index built by RDF-3X and Jena TDB were 87 GB and 110 GB,
respectively.  \riq's filtering index was 16 GB in size and had 2620
unions. Note that the size of the LUBM and BTC 2012 datasets were 217
GB and 218 GB, respectively. When constructing the filters of both
datasets in \riq, we set the false positive rate $\epsilon$ equal to
5\%.

\subsection{Query Processing}

We measured the wall-clock time taken to process each query in both
cold and warm cache settings, and report the average over 3 runs in
Table~\ref{table-queries}. Jena TDB was executed with its default
statistics-based optimization.

For LUBM, \riq{} processed queries with large, complex BGPs (L1-L3)
significantly faster than RDF-3X and Jena TDB in both cold and warm
cache settings. For BTC 2012, \riq{} was significantly faster than
RDF-3X in processing queries B1 and B2. This demonstrates that the {\em
  decrease-and-conquer} approach of \riq{} is more effective than the
popular join-based processing (by first matching individual triple
patterns) on queries with large, complex BGPs. All of the
large, complex queries had at least one undirected cycle. \riq{}
identified a maximum of 22 candidate groups for queries L1-L3 and 4
candidate groups for queries B1 and B2.

Next, we report the performance of \riq{} on queries with (small) BGPs
containing less than 8 triple patterns (L4-L12 and
B3-B7). Interestingly, on LUBM, \riq{} was faster than RDF-3X and Jena
TDB for six out of the nine queries in both cold and warm cache
settings. On BTC 2012, \riq{} was the fastest in the cold cache
setting for three out of the five queries. However, RDF-3X was the
fastest in the warm cache setting for four out of the five
queries. Finally, we compared the three approaches based on the
geometric mean of their query processing times. Clearly, \riq{} was
the winner for both LUBM and BTC 2012.



\begin{table*}[tbh]
\begin{center}
\begin{tabular}{|c|c|c|c|c|c|c|c||c|c|c|}
\hline
\multicolumn{1}{|c|}{\bf{\small{Dataset}}} &
\multicolumn{1}{c|}{\bf{\small{Query}}} &
\multicolumn{1}{c|}{\bf{\small{Type}}} &
\multicolumn{1}{c|}{{\bf \small{\# of}}} &
\multicolumn{1}{c|}{{\bf \small{\# of}}} &
\multicolumn{3}{c|}{{\bf \small{Cold cache}}} &
\multicolumn{3}{c|}{{\bf \small{Warm cache}}}\\
\multicolumn{1}{|c|}{} &
\multicolumn{1}{c|}{} &
\multicolumn{1}{c|}{} &
\multicolumn{1}{c|}{{\bf \small{triple}}} &
\multicolumn{1}{c|}{{\bf \small{results}}} &
\multicolumn{3}{c|}{Time taken (in secs)} &
\multicolumn{3}{c|}{Time taken (in secs)}\\ \cline{6-11}
\multicolumn{1}{|c|}{} &
\multicolumn{1}{c|}{} &
\multicolumn{1}{c|}{} &
\multicolumn{1}{c|}{\bf{\small{patterns}}} &
\multicolumn{1}{c|}{} &
\multicolumn{1}{c|}{\bf{\small{\riq{}}}} &
\multicolumn{1}{c|}{\bf{\small{RDF-3X}}} &
\multicolumn{1}{c|}{\bf{\small{Jena TDB}}} &
\multicolumn{1}{c|}{\bf{\small{\riq{}}}} &
\multicolumn{1}{c|}{\bf{\small{RDF-3X}}} &
\multicolumn{1}{c|}{\bf{\small{Jena TDB}}}\\
\hline
\hline
\multirow{12}{*}{LUBM} & \small{L1} & \small{large} & \small{18} & \small{24} & \cellcolor{Gray}\bf{\small{28.09}} & \small{73.65} & \small{263.31} & \cellcolor{Gray}\bf{\small{2.8}} & \small{63.93} & \small{4.22}\\
\hhline{|~|-|-|-|-|-|-|-|-|-|-|}
& \small{L2} & \small{large} & \small{11} & \small{7,082} & \cellcolor{Gray}\bf{\small{300.08}} & \small{$77,315^\dagger$} & \small{77,315} & \cellcolor{Gray}\bf{\small{49.03}} & \small{$64,637^\dagger$} & \small{64,637}\\
\hhline{|~|-|-|-|-|-|-|-|-|-|-|}
& \small{L3} & \small{large} & \small{22} & \small{0} & \cellcolor{Gray}\bf{\small{24.59}} & \small{1692.63} & \small{179.19} & \cellcolor{Gray}\bf{\small{0.39}} & \small{1688.94} & \small{1.97}\\
\hhline{|~|-|-|-|-|-|-|-|-|-|-|}
& \small{L4} & \small{small} & \small{6} & \small{2,462} & \cellcolor{Gray}\bf{\small{229.95}} & \small{1986.21} & \small{698.08} & \cellcolor{Gray}\bf{\small{27.46}} & \small{1899.1} & \small{664.75}\\
\hhline{|~|-|-|-|-|-|-|-|-|-|-|}
& \small{L5} & \small{small} & \small{1} & \small{25,205,352} & \cellcolor{Gray}\bf{\small{576.96}} & \small{995.26} & \small{1130.43} & \cellcolor{Gray}\bf{\small{567.2}} & \small{948.53} & \small{1127.37}\\
\hhline{|~|-|-|-|-|-|-|-|-|-|-|}
& \small{L6} & \small{small} & \small{6} & \small{468,047} & \cellcolor{Gray}\bf{\small{506.93}} & \small{888.84} & \small{1119.31} & \cellcolor{Gray}\bf{\small{489.36}} & \small{847.59} & \small{1144.11}\\
\hhline{|~|-|-|-|-|-|-|-|-|-|-|}
& \small{L7} & \small{small} & \small{1} & \small{79,163,972} & \cellcolor{Gray}\bf{\small{892.7}} & \small{1215.53} & \small{aborted} & \cellcolor{Gray}\bf{\small{871.12}} & \small{1153.31} & \small{aborted}\\
\hhline{|~|-|-|-|-|-|-|-|-|-|-|}
& \small{L8} & \small{small} & \small{2} & \small{10,798,091} & \cellcolor{Gray}\bf{\small{507.43}} & \small{805.41} & \small{1346.17} & \small{497.69} & \cellcolor{Gray}\bf{\small{70.35}} & \small{1395.48}\\
\hhline{|~|-|-|-|-|-|-|-|-|-|-|}
& \small{L9} & \small{small} & \small{6} & \small{440,834} & \cellcolor{Gray}\bf{\small{538.99}} & \small{979.79} & \small{1137.38} & \cellcolor{Gray}\bf{\small{519.22}} & \small{947.07} & \small{1142.73}\\
\hhline{|~|-|-|-|-|-|-|-|-|-|-|}
& \small{L10} & \small{small} & \small{5} & \small{8,341} & \small{18.72} & \small{11.11} & \cellcolor{Gray}\bf{\small{7.15}} & \cellcolor{Gray}\bf{\small{0.51}} & \small{6.39} & \small{3.19}\\
\hhline{|~|-|-|-|-|-|-|-|-|-|-|}
& \small{L11} & \small{small} & \small{4} & \small{172} & \small{12.19} & \cellcolor{Gray}\bf{\small{1.98}} & \small{5.79} & \small{0.41} & \cellcolor{Gray}\bf{\small{0.25}} & \small{1.13}\\
\hhline{|~|-|-|-|-|-|-|-|-|-|-|}
& \small{L12} & \small{small} & \small{6} & \small{0} & \small{103.14} & \cellcolor{Gray}\bf{\small{22.33}} & \small{725.93} & \small{26.76} & \cellcolor{Gray}\bf{\small{19.83}} & \small{703.26}\\
\hline
\multicolumn{5}{|c|}{\small{Geometric Mean (excludes L7 for Jena TDB)}} & \cellcolor{Gray}\bf{\small{144.09}} & \small{375.98} &\small{448.65} & \cellcolor{Gray}\bf{\small{29.89}} & \small{233.25} & \small{160.83}\\
\hline
\hline
\multirow{12}{*}[2em]{BTC-2012} & \small{B1} & \small{large} & \small{19} & \small{6} & \cellcolor{Gray}\bf{\small{8.81}} & \small{1560.12} & \small{16.16} & \cellcolor{Gray}\bf{\small{1.19}} & \small{1497.74} & \small{13.14}\\
\hhline{|~|-|-|-|-|-|-|-|-|-|-|}
& \small{B2} & \small{large} & \small{21} & \small{5} & \cellcolor{Gray}\bf{\small{14.56}} & \small{364.93} & \small{19.34} & \cellcolor{Gray}\bf{\small{6.52}} & \small{362.49} & \small{16.86}\\
\hhline{|~|-|-|-|-|-|-|-|-|-|-|}
& \small{B3} & \small{small} & \small{4} & \small{47,493} & \cellcolor{Gray}\bf{\small{41.01}} & \small{56.42} & \small{373.59} & \small{1.83} & \cellcolor{Gray}\bf{\small{0.82}} & \small{20.13}\\
\hhline{|~|-|-|-|-|-|-|-|-|-|-|}
& \small{B4} & \small{small} & \small{6} & \small{146,012} & \cellcolor{Gray}\bf{\small{42.17}} & \small{48.55} & \small{321.56} & \small{3.59} & \cellcolor{Gray}\bf{\small{2.37}} & \small{35.99}\\
\hhline{|~|-|-|-|-|-|-|-|-|-|-|}
& \small{B5} & \small{small} & \small{7} & \small{1,460,748} & \cellcolor{Gray}\bf{\small{70.15}} & \small{74.86} & \small{3541.99} & \small{32.38} & \cellcolor{Gray}\bf{\small{28.64}} & \small{3540.28}\\
\hhline{|~|-|-|-|-|-|-|-|-|-|-|}
& \small{B6} & \small{small} & \small{5} & \small{0} & \small{20.39} & \small{$40,140^\dagger$} & \cellcolor{Gray}\bf{\small{14.89}} & \cellcolor{Gray}\bf{\small{0.64}} & \small{$40,140^\dagger$} & \small{12.83}\\
\hhline{|~|-|-|-|-|-|-|-|-|-|-|}
& \small{B7} & \small{small} & \small{5} & \small{12,101,709} & \small{221.86} & \cellcolor{Gray}\bf{\small{210.37}} & \small{1925.27} & \small{184.86} & \cellcolor{Gray}\bf{\small{118.84}} & \small{1817.85}\\
\hline
\multicolumn{5}{|c|}{\small{Geometric Mean}} & \cellcolor{Gray}\bf{\small{35.45}} & \small{372} & \small{168.22} & \cellcolor{Gray}\bf{\small{5.7}} & \small{105.36} & \small{74.92}\\
\hline
\end{tabular}
\vspace*{-2ex}
\caption{Query processing times (in seconds) for LUBM and
  BTC-2012. Best times are shown in bold within shaded
  cells. $X^\dagger$ indicates that the query ran for more than $X$
  seconds and was terminated.}
\label{table-queries}
\vspace*{-4ex}
\end{center}
\end{table*}

\vspace*{-1ex}
\section{Conclusions}
We presented \riq, a new approach {} for indexing large RDF datasets
containing quads. \riq{} employs a {\em decrease-and-conquer} approach
to efficiently process SPARQL queries. Through our experiments, we
demonstrate that \riq{} enables efficient SPARQL query processing on
large RDF datasets with more than a billion quads.

\comment{
It significantly outperformed popular tools like RDF-3X and Jena TDB
for queries with large, complex BGPs. It also achieved competitive
performance for queries with small BGPs.
}

\subsubsection*{Acknowledgement}
This work was supported by the National Science Foundation under Grant
No. 1115871.

\vspace*{-2ex}
\begin{small}
\bibliographystyle{abbrv}
\bibliography{rdf}
\end{small}
\end{document}